\documentclass[apj]{emulateapj} \usepackage{apjfonts} 
\usepackage{graphicx}

\shorttitle{The Third VLBA Calibrator Survey}
\shortauthors{Petrov et al.}
\received{September 28, 2004}
\revised{October 14, 2004}
\accepted{October 20, 2004}

\journalinfo{Astronomical journal}

\bibpunct{(}{)}{,}{a}{}{,}
\makeindex
\citeindextrue

\newcommand{\getlength}[1]{\ifx#1\end \let\next=\relax
            \else\advance\count255 by1 \let\next=\getlength\fi \next}
\newcount\switch
\newcommand{\Begmat}{\ifm\switch=1\else\switch=0$\fi}
\newcommand{\Endmat}{\ifnum\switch=0$\fi}
\newcommand{\ifnularg}[1]{ \count255=0 \getlength#1\end \ifnum\count255=0 }
\newcommand{\ifm}{\makebox{}\ifmmode}
\long\def\ifundefined#1#2#3{\expandafter\ifx\csname
  #1\endcsname\relax#2\else#3\fi}
\newcommand{\beq}   { \begin{eqnarray} }
\newcommand{\eeq}[1]{ \ifnularg{#1} end{eanarray} \else
                      \label{#1}\end{eqnarray}    \fi }
\newcommand{\flo}[2]{ {\Begmat
            \ifnularg{#1} 10^{#2}\else \mbox{#1} \cdot 10^{#2}\fi\Endmat}}
\newcommand{\eref}[1]{(\ref{#1})}
\begin{document}

\title{The Third VLBA Calibrator Survey --- VCS3}
\author{L. Petrov}
\affil{NVI, Inc./NASA GSFC, Code 926, Greenbelt, MD 20771, USA}
\email{Leonid.Petrov@lpetrov.net}
\author{Y. Y. Kovalev\altaffilmark{\dag}}
\affil{National Radio Astronomy Observatory,
       P.O.\ Box 2, Green Bank, WV 24944, USA;}
\affil{Astro Space Center of P.N. Lebedev Physical Institute,
       Profsoyuznaya 84/32, 117997 Moscow, Russia}
\email{ykovalev@nrao.edu}
\author{E. Fomalont}
\affil{National Radio Astronomy Observatory,
       520 Edgemont Road, Charlottesville, VA~22903--2475, USA}
\email{efomalon@nrao.edu}
\author{D. Gordon}
\affil{Raytheon/NASA GSFC, Code 926, Greenbelt, MD 20771, USA}
\email{dgg@leo.gsfc.nasa.gov}
\altaffiltext{\dag}{Jansky Fellow, National Radio Astronomy Observatory}

\begin{abstract}

  This paper presents the third extension to the Very Large Baseline Array
(VLBA) Calibrator Survey, containing 360 new sources not previously
observed with very long baseline interferometry (VLBI). The survey, based on 
three 24 hour VLBA observing sessions, fills the areas on the sky above 
declination $-45\degr\ $ where the calibrator density is less than 
one source within a 4\degr\ radius disk at any given direction. 
The positions were derived from astrometric analysis of the group delays 
determined at 2.3 and 8.6~GHz frequency bands using the Calc/Solve 
software package. The VCS3 catalogue of source positions, plots of 
correlated flux density versus the length of projected baseline, contour 
plots and fits files of naturally weighted CLEAN images as well as 
calibrated visibility function files are available on the Web at 
\url{http://gemini.gsfc.nasa.gov/vcs3}

\end{abstract}


\section{Introduction}

   Catalogues of positions of compact radio sources with the highest
precision are the basis for many applications. Among of them are
imaging faint radio sources in phase referencing mode, monitoring the
Earth's rotation, space geodesy, and space navigation. The method of VLBI
first proposed by \cite{mat65} allows us to determine positions of
sources with nanoradian precision (1~nrad $\approx$~0.2 mas). The
first catalogue of source positions determined with VLBI
\citep{first-cat} contained only 35 sources. It is desirable for the
above mentioned applications to have many more sources. In
1998 the ICRF catalogue of 667 sources produced by analyzing VLBI
observations made in the framework of space geodesy programs was
published \citep{icrf98}. Six years later, the extension of this
catalogue, ICRF-Ext2, with 109 more sources, was published
\citep{icrf-ext2-2004}. However, the sky coverage of these catalogues
was still not sufficient for some applications. Successful phase 
referencing requires a calibrator within 4\degr\ from the 
target source with a precise position and known source structure.
The probability of finding a calibrator from the ICRF-Ext2
catalogue within 4\degr\ of any random target above $ -45\degr\ $ 
declination is about 50\%. Also, the ICRF catalogue does not provide 
source maps. In order to overcome these deficiencies, twelve 24 hour 
observing sessions with the VLBA, called VLBA Calibrator Surveys, were 
performed. Analysis of these observations produced the VCS1 \citep{vcs1} 
and VCS2 \citep{vcs2} catalogues of 1611 new sources not listed in the ICRF
catalogue. Approximately 80\% of these sources are suitable as phase
referencing calibrators and as target sources for space geodesy
programs. 

  Combining all suitable calibrators from the ICRF-Ext2, VCS1 and VCS2
catalogues we get a list of 1986 sources. Over 89.5\% of the sky north of
$-45\degr\ $ declination we find at least one calibrator within a 4\degr\ 
radius disk at any given direction. In this paper we present an extension 
of the VCS1 and VCS2 catalogues, called the VCS3 catalogue, mainly 
concentrating on the other 10.5\% of the sky where the source density 
is lowest. Since the observations, calibrations, astrometric solutions and 
imaging are similar to that of VCS1, most of the details are described 
by \cite{vcs1}. In section 2 we describe the strategy for selecting the 
450 candidate sources observed in three 24 hour sessions with the VLBA.
In section 3 we briefly outline the observations and data analysis. We
present the catalogue in section 4, and summarize our results in section 5.

\section{Source selection}

   Having unlimited resources one could try to observe all sources stronger 
than some limiting flux. However, only sources with bright compact components 
can be detected with VLBI and might be useful for phase referencing or 
geodetic applications. \citet{kellermann} first showed that the distribution 
of sources over spectral index $\alpha \enskip (F \propto \nu^\alpha)$ has 
two peaks: one near $\alpha=-1$ (steep spectrum) and another near $\alpha=0$
(flat spectrum). Later, it was confirmed  that extended objects dominate 
in the steep spectrum population \citep[see the review by][]{ko88}. 
In compact regions, which have the synchrotron mechanism of emission, 
the peak in the spectrum caused by synchrotron self-absorption has 
frequencies higher than 1~GHz due to their small size. Thus, if a dominating 
emission comes from compact regions, the spectrum of the total flux density 
will be predominantly flat or inverted.

   Examining spectral indexes of the PKScat90 catalogue \citep{pkscat90a}, 
which is believed to be complete at the 95\% level for sources with flux 
density \mbox{$F > 0.25$~Jy} at 2.7~GHz \citep{pkscat90}, we found that 
only 26\% of sources have spectral indexes \mbox{$ \alpha > -0.5$}. 
\citet{class} used NVSS \citep{nvss} and GB6 \citep{gb6} catalogues for 
selection of sources with spectral indexes \mbox{$> -0.5$} at 5~GHz, and 
they found 11685 sources brighter than 30~mJy, or 24\% of the total 
number of sources. It should be noted that a significant fraction of the 
sources with spectral index in the range \mbox{[--0.5, --0.4]} belongs to 
the steep spectrum population and are expected to have extended structure. 
Thus, taking sources at random, the probability that a source will be 
compact, and therefore, detectable with VLBI does not exceed 20\%. 
A carefully designed strategy for source selection can significantly 
increase the yield of detections.

   We considered all bright sources at X-band (8.6~GHz) and S-band (2.3~GHz) 
with spectral indexes greater than $-0.5$ as candidates for the VCS3 
campaign. In order to find these sources, we first analyzed catalogues 
of compact flat-spectrum sources: JVAS \citep{jvas}, CLASS \citep{class}, 
Parkes quarter-Jansky catalogue \citep{parkes-qJy}, VLA 
South \citep{phcal-south} and selected all sources brighter than 100~mJy at 
declinations above $-45\degr\ $ with arc distances from any known calibrator 
greater than 3.9\degr. This produced a list of 374 sources. Not all these 
catalogues provide flux density at more than one band. We used the CATS 
database \citep{cats} for cross-checking the reported flux density and for 
acquiring information about flux measurements for these objects from other 
sources. Unfortunately, most of these measurements were done at different 
epochs. Since many flat spectrum sources are variable, the estimates of the 
spectral index made at different epochs may be unreliable. Instantaneous 
1--22~GHz broad-band spectra at 4--6 frequencies for 3000~AGN with flux 
density greater than 100~mJy at 2.3~GHz and \mbox{$\alpha>-0.5$}, 
including many sources from this list, were observed with the 600 meter 
ring radio telescope of the Russian Academy of Sciences \mbox{RATAN--600} 
in transit mode \citep[see method of observations and data reduction 
in][]{Kovalev_etal99} as a part of an ongoing survey program. All flat 
spectrum sources from this sample with flux densities greater than 
200~mJy at X and/or S band and not previously observed with VLBI, a total 
of 116 sources, were included in the VCS3 candidates list. 
After cross checking we removed sources from the list for which the estimate 
of the spectral index was either unreliable or less than $-0.5$. The total 
number of sources selected from these catalogues was 300.

   After picking up these 300 sources from all known compact flat-spectrum 
catalogues we still had areas where the density of calibrators 
was less than the targeted density. In order to fill these areas we used the 
NVSS catalogue of about 1.8 million sources \citep{nvss}. First, 
we selected 2704 sources with total flux density greater than 100~mJy at 
1.4~GHz in the areas where the distance to the closest calibrator and 
already selected candidates was greater than 4\degr. The NVSS catalogue 
does not provide information about spectral index, so we checked 
these 2704 sources against the CATS database, which nowadays contains one of 
the most complete collections of radio catalogues, and found the sources 
observed under other campaigns at different frequency bands. If they were 
not found in any other catalogue, for example, they were absent in the GB6 
catalogue with flux limit 18~mJy and in the PMN \citep{pmn} catalogue with 
flux limit 40--70~mJy at 5~GHz, this indicates that with a high probability 
their spectral index is significantly less than $-0.5$. In total, this 
initial list had 302 sources.

   We examined the spectrum provided by CATS and subjectively ascribed a class 
to each source:
\begin{itemize} 

  \item[0 ---] the source has flat spectrum, and the estimate of the spectral 
               index is reliable (e.g., \mbox{RATAN--600} instantaneous 
               spectrum is available), and the estimated correlated flux 
               density is greater than 100~mJy;

  \item[1 ---] the estimate of the spectral index is less reliable, 
               or the spectral index is in the range [--0.4, --0.2],
               and the estimated correlated flux density is greater than 
               100~mJy;

  \item[2 ---] the estimate of the spectral index is unreliable or in the range 
               of [--0.5, --0.4], and/or the estimated correlated flux density 
               is less than 100~mJy.
\end{itemize} 

   Then for each source we computed a score based on angular distance to 
the closest calibrator, flux density, spectral index and class:

\beq
   S \; = \; 10^{3} \, (( D - D_{lim})/D_{lim} )^3 \; + \;
          10^{5} \, (\alpha+0.5)^3 \; + \nonumber \\
          \flo{2}{2} \, (2-C) \; + \; \lg F
\eeq{e:e1}

  where $D$ --- angular distance to the closest calibrator or already selected
candidate, $D_{lim}$ minimal angular distance (4\degr), $\alpha$ --- spectral 
index, $C$ --- class, $F$ --- flux density in Jy extrapolated to 8.6~GHz. 
An iterative procedure computed the score, selected the source with the 
largest score and ran again. Using this method we selected 150 sources with 
the highest scores in addition to the 300 sources previously chosen.

\section{Observations and data analysis}

  The VCS3 observations were carried out in three 24 hour observing
sessions with the VLBA on 2004 April 30, 2004 May 08, and 2004 May 27.
Since the correlated flux density for selected candidates was expected 
to be different, we split our sources into three categories:
\begin{itemize}
   \item[{\sf a} ---] sources of class 0 brighter than 200~mJy;
   \item[{\sf b} ---] sources of class 0 with flux density in the range 
                       100--200~mJy or sources of class 1 brighter than 200~mJy;
   \item[{\sf c} ---] all other sources.
\end{itemize}

  Integration time was chosen to be 90 seconds for sources from category 
{\sf a}, 180 seconds from category {\sf b}, and 300 seconds from category 
{\sf c}. The sources above $-20$\degr\ declination were scheduled for two 
scans at different hour angles in order to improve the $uv$ coverage, while
the sources in \mbox{[$-45\degr, \-20\degr$]} declination range were
scheduled only for one scan. The target sources were observed in a 
sequence designed to minimize loss of time from antenna slewing. In 
addition to the target sources, 77 strong sources were taken from the 
GSFC astrometric and geodetic catalogue \mbox{2004a\_astro}
\footnote{\url{http://gemini.gsfc.nasa.gov/solutions/astro}}. 
Observations of 3--4 strong sources from this list were made every 1--1.5 
hours during observing sessions. These observations were scheduled in such 
a way, that at each station at least one of these sources was observed at 
an elevation angle less than 20\degr, and at least one at an elevation 
angle greater than 50\degr. The purpose of these observations was to provide
calibration for mismodelled troposphere path delays and to tie the VCS3 
source positions to the ICRF catalogue. On average, the antennas were 
on-source 70\% of the time.

  The observations used the VLBA dual-frequency geodetic mode, observing
simultaneously at 2.3~GHz and 8.6~GHz. Each band was separated into four 
8 MHz channels (IFs) which spanned 140 MHz at 2.3~GHz and 490 MHz at 8.6~GHz, 
in order to provide precise measurements of group delays for astrometric 
processing. Since the a~priori coordinates of many candidates were 
expected to have errors of up to 30\arcsec, the data were correlated with 
an accumulation period of 1~second in 64~frequency channels in order 
to provide extra-wide windows for fringe searching.

  Processing of the VLBA correlator output to obtain the astrometric 
observables was made using the Astronomical Image Processing System (AIPS)
\citep{aips}. First, phases of the calibration signals injected into the
receivers were subtracted from the fringe phases for each scan and each 
channel. Then the phases of a single strong calibrator scan were subtracted 
from each scan to flatten out the phases across the band of each individual 
channel. Sampler bias correction together with initial amplitude calibration 
was performed using the recorded system temperatures and gain tables 
(tasks ``ACCOR'' and ``APCAL''). Task ``FRING'' was then run for each 
baseline separately (with SNR cutoff~=~3) to obtain the residual 
narrow-band group delays for each channel, residual single channel fringe 
phases, and residual single channel phase delay rates. Task ``MBDLY'' was 
then used to compute the residual group delays from the 4 single channel 
fringe phases at each frequency band. Then task ``CL2HF'' was used to compute 
the total group delays, phase delay rates and fringe phases by combining 
the residuals with the correlator model, and converting them from geocenter 
to reference station quantities. Finally, the AIPS output was reformatted for 
input into the Calc/Solve geodetic/astrometric VLBI analysis software 
package\footnote{\url{http://gemini.gsfc.nasa.gov/solve}} using task 
``HF2SV''.

  The software package Calc/Solve was used for determination of accurate 
positions of all detected sources. Astrometric analysis was performed in 
several steps. First, we used narrow-band group delays determined over each 
8~MHz wide channel and incoherently averaged separately over all channels 
within X and S band. Narrow-band group delays are less precise than 
wide-band group delays, but have very wide ambiguity spacings: 4~mks. In 
the first step X-band single-band delay data (and S-band data for the 
sources detected only at S-band) were used for estimation of new source 
positions and clock functions modeled by second degree polynomials. These 
position estimates were used in an S-band wide-band group delay solution 
for short baselines between stations Pietown, Kitt Peak, Owens Valley, 
Los Alamos and Fort Davis in order to resolve the group delay ambiguities 
with spacings of 100~ns and then again solve for source positions in the 
second step. In step three these estimates were used in an S-band group 
delay solution for longer baselines for further 
ambiguity resolution. In the fourth step the previous estimates of source 
positions were used for ambiguity resolution of X-band group delays 
(with spacings 29~ns) on short baselines, and a new solution was made.  
These estimates then were used for resolving X-band group delays at long 
baselines in the next step. In the sixth step ionosphere free combinations 
of X and S band wide-band group delays were used for data analysis. 
The estimation model at this step included coefficients of linear splines
for clocks of all stations, except a reference station, and troposphere 
zenith path delays at all sites. The time interval for linear splines was 
60~minutes. The data were carefully analyzed for the presence of any 
remaining errors in group delay ambiguity resolution and for outliers. 
About 1\% of the data were deselected, primarily because the fringe fitting 
algorithm picked a secondary maximum of the delay resolution function. If 
only two observations of a source are used in a solution, errors in group 
delay ambiguity resolution cannot be detected. However, if more observations 
are used in a solution, errors in group delay ambiguity resolution can 
result in a significant misfit. Therefore, analysis of observations of 
sources with numerous misfits allowed us to detect group delay ambiguity 
resolution errors with at least 3--6 observations. Being conservative, we 
set a minimum number of 8 observations needed to rule out the possibility 
of group delay ambiguity resolution errors. After outlier elimination 
we determined additive baseline-dependent corrections to the a~priori 
weights in such a manner that the ratios of the sums of weighted squares 
of post-fit residuals to their mathematical expectation were close to unity.

  In the final solution all available astrometric and geodetic data from 
August 1979 to August 2004, including the three VCS3 sessions, were used in 
a single least square solution. Estimated parameters included positions 
and velocities of stations, coordinates of all sources, Earth orientation
parameters, clock functions and troposphere zenith path delays. The only 
differences in treatment of the VCS3 sessions with respect to other 
experiments were that we did not estimate troposphere gradients and the 
time interval for clocks and troposphere zenith path delays were 60 minutes 
instead of 20 minutes. Since a few sources were detected in only one band,
we made three solutions: using 1) ionosphere free linear combinations of 
X-band and S-band wide-band group delays, 2) only X-band group delays 
and 3) only S-band group delays.

  Since the equation for time delay is invariant with respect to the 
group of coordinate transformations, observations themselves do not determine
source positions, but only a family of positions. In order to pick a specific
element of this family, we imposed boundary conditions by applying net-rotation 
constraints on the positions of 212 sources listed as ``defined'' in the ICRF 
catalogue such that the set of new positions of these 212 sources did not 
have a net-rotation with respect to the set of old position from the ICRF. 
More details about this procedure can be found in \citet{vcs1}.

\begin{deluxetable*}{
                    r @{\hspace{0.25em}} r  
                    c c c @{\hspace{1.5em}}c  
                    c c c @{\hspace{1.5em}}c  
                    c c c
                   }
\tablecaption{
              Additive corrections to formal uncertainties applied for
              derivation of errors of source position in VCS3 
              catalogue \label{t:err} 
             }
\tabletypesize{\scriptsize}
\tablehead{
           \multicolumn{2}{c}{Declination} &
           \multicolumn{3}{c}{X/S }        &&
           \multicolumn{3}{c}{X-band}      &&
           \multicolumn{3}{c}{S-band} 
           \\
	   \cline{3-5}
	   \cline{7-9}
	   \cline{11-13}
           \multicolumn{2}{c}{zone}                & 
           \colhead{\# pts}                        &
           \colhead{$E_\alpha$}                    &
           \colhead{$E_\delta$}                    &&
           \colhead{\# pts}                        &
           \colhead{$E_\alpha$}                    &
           \colhead{$E_\delta$}                    &&
           \colhead{\# pts}                        &
           \colhead{$E_\alpha$}                    &
           \colhead{$E_\delta$}                    
           \\
           \multicolumn{2}{c}{(deg)} &&
           \colhead{(mas)} & 
           \colhead{(mas)} && 
           \colhead{ }     & 
           \colhead{(mas)} & 
           \colhead{(mas)} && 
           \colhead{ }     & 
           \colhead{(mas)} & 
           \colhead{(mas)} 
          }
\startdata
 ( +20.0, & +90.0 ) & 33 & 0.2 & 0.4 &&  96 & 0.6 & 0.5 && 94 &  5.1 & 4.7  \\
 ( \hspace{1.0em}0.0, & +20.0 ) & 16 & 0.3 & 0.2 &&  77 & 0.5 & 0.6 && 77 &  4.2 & 7.2  \\
 ( --20.0, &   0.0 ) & 11 & 0.2 & 0.3 &&  64 & 0.8 & 1.5 && 64 &  7.0 & 16.8 \\
 ( --30.0, & -20.0 ) & 11 & 0.2 & 0.7 &&  44 & 1.2 & 1.8 && 44 & 13.0 & 27.6 \\
 ( --45.0, & -30.0 ) & 13 & 0.3 & 0.4 &&  76 & 2.4 & 7.3 && 75 & 40.0 & 40.0  
\enddata
\end{deluxetable*}

\begin{figure}[b]
   \begin{center}
   \resizebox{0.9\hsize}{!}{\includegraphics{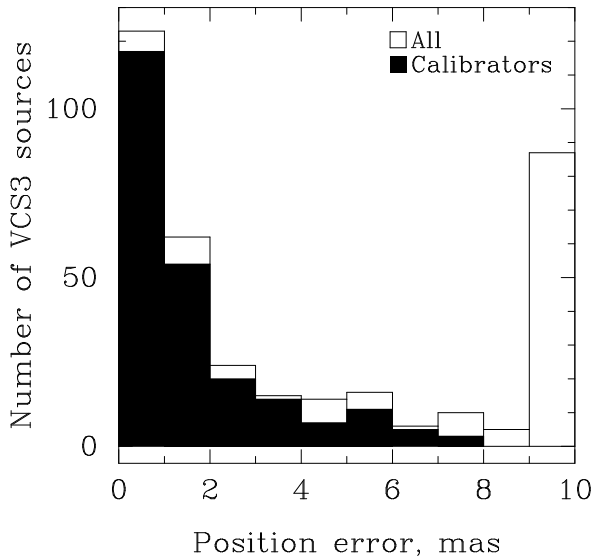}}
   \end{center}
   \par\vspace{-3ex}\par
   \caption{Histogram of semi-major error ellipse of position errors.
            The last bin shows errors exceeding 9~mas. 
            \label{f:hist}
           }
\end{figure}

  In order to evaluate position errors we made a special solution using only
the three VCS3 sessions in which we estimated coordinates of all sources and 
applied no-net-rotation constraints on 20 sources common in the VCS3 and 
ICRF defining lists. Since 77 sources used in the VCS3 campaign were 
intensively observed earlier under various geodetic and astrometric programs, 
we used the differences between very accurate positions derived from analysis 
of these observations and positions from the VCS3 sessions as a measure of 
the upper limit of VCS3 catalogue errors. We represent estimates of errors 
in right ascension ($\sigma_\alpha$) and in declination ($\sigma_\delta$) 
of the VCS3 catalogue in the form:
\beq
    \begin{array}{l}
       \sigma_\alpha = \sqrt{ (1.5 \cdot \sigma^r_\alpha)^2 + 
                         (E_\alpha(\delta) / cos\ \delta)^2  }
                                                              \vspace{2.5ex} \\
       \sigma_\delta = \sqrt{ (1.5 \cdot \sigma^r_\delta)^2 + 
                                E^2_\delta(\delta) }
   \end{array}
\eeq{e:e2}

  where $\sigma_\alpha^r$ and $\sigma_\delta^r$ stand for the formal 
uncertainties derived by the errors propagation law from fringe amplitude 
signal to noise ratios, $E_\alpha$ and $E_\delta$ are additive reweighting 
parameters. The multiplicative factor 1.5 in \eref{e:e2}, first found 
by \citet{errfactor}, accounts for underestimated systematic errors in the 
computation of uncertainties of wide-band group delays and affects 
uncertainties of estimates of all parameters from VLBI, including source 
coordinates. We split the sources into five declination zones, and for 
each group we found the values of $E_\alpha$ and $E_\delta$ which made 
ratios of the weighted sum of squares of residuals to their mathematical 
expectation close to unity. Using this techniques we computed additive 
corrections to uncertainties of the estimates of X-band only and S-band 
only solutions by analyzing the differences between dual band and single 
band solutions. Reweighting parameters are presented in Table~\ref{t:err}. 
The histogram of source position errors is presented in Figure~\ref{f:hist}.

  For imaging purposes the data were re-processed with the global fringe
fitting procedure (task ``FRING'' ) with SNR cutoff~4 at each individual 
channel. Global, station-based fringing allows detection of weaker sources 
than the baseline-based fringing used for astrometric analysis. 
(We did not use global fringing for astrometric analysis, since as we 
found earlier, it gives less reliable group delays than baseline-based 
fringing). A bandpass correction was then determined with the task ``BPASS''. 
After initial editing the fringe amplitude and phases were averaged across 
the channels. We used the AIPS task ``CALIB'' in order to perform the first
step of the self-calibration procedure applying the point source model. 
The whole scan length was used to find the combined solution for the channels 
with the SNR cutoff~3. After that we imported data to the Caltech DIFMAP 
package~\citep{difmap}. We have adapted the automated procedure 
originally written by Greg Taylor~\citep{difmap-script} for 
self-calibration and imaging of the VCS3 short snapshot observations.

For each detected source at each band, we computed two estimates of 
correlated flux density: one averaged over the entire image (total CLEAN 
flux density) and another averaged only over the spatial frequencies 
$> 45 M\lambda$ at S-band and and $> 170 M\lambda$ at X-band, which
corresponds to the range \mbox{[0.7, 1.0]} of the maximum spatial 
frequencies of the VLBA. The latter estimate characterizes the flux density 
of unresolved components. Errors of our estimates of total correlated flux 
density of sources brighter than 100~mJy are determined mainly by the 
accuracy of amplitude calibration, which for the VLBA, according 
to \citet{VLBA_summ}, is at the level of 5\% at 1--10~GHz. This estimate 
was confirmed by the comparison of the correlated flux density with the 
single-dish flux density measured with \mbox{RATAN--600} in October 2003 
for several slowly varying sources without extended structure. The 
contribution of fringe amplitude errors is significant for sources with 
flux density \mbox{$< 100$~mJy}. The error of correlated flux density for 
this group of sources is about 10\%. It should be noted that in the case 
of a source with asymmetric extended structure the estimate of the 
correlated flux density from unresolved components may be biased due 
to a significant non-uniformity of $uv$ coverage at long spacings.

\begin{deluxetable*}{ c l l l r r r r r r r r r l}
\tablecaption{The VCS3 catalogue \label{t:cat}}
\tabletypesize{\scriptsize}
\tablehead{
   \colhead{} & 
   \multicolumn{2}{c}{Source name}                   &
   \multicolumn{2}{c}{J2000.0 Coordinates}           &
   \multicolumn{3}{c}{Errors (mas)}                  &
   \colhead{}                                        & 
   \multicolumn{4}{c}{Correlated flux density (Jy)}  &
   \colhead{} 
   \\
   \multicolumn{9}{c}{}         &
   \multicolumn{2}{c}{8.6 GHz}  &
   \multicolumn{2}{c}{2.3 GHz}  \\
   \\
   \colhead{Class}    &
   \colhead{IVS}      &
   \colhead{IAU}      &
   \colhead{Right ascension} &
   \colhead{Declination}     &
   \colhead{$\Delta \alpha$} &
   \colhead{$\Delta \delta$} &
   \colhead{Corr}   &
   \colhead{\# Obs} &
   \colhead{Total } &
   \colhead{Unres } &
   \colhead{Total } &
   \colhead{Unres } &
   \colhead{Band}
   \\
   \colhead{(1)}    &
   \colhead{(2)}    &
   \colhead{(3)}    &
   \colhead{(4)}    &
   \colhead{(5)}    &
   \colhead{(6)}    &
   \colhead{(7)}    &
   \colhead{(8)}    &
   \colhead{(9)}    &
   \colhead{(10)}   &
   \colhead{(11)}   &
   \colhead{(12)}   &
   \colhead{(13)}   &
   \colhead{(14)} 
   }
\startdata
C  & 0001-120 & J0004-1148 & 00 04 04.914997 & -11 48 58.38564 &    0.24  &  0.40  &  -0.069 &     20 &$  0.62 $ & \nodata  & $ 0.76 $ & \nodata   &  X/S \\
-- & 0003+340 & J0006+3422 & 00 06 07.382435 & +34 22 20.41138 &    1.83  &  4.46  &  -0.233 &      5 &$  0.06 $ & \nodata  & \nodata  & \nodata   &  X   \\
C  & 0009+655 & J0012+6551 & 00 12 37.671095 & +65 51 10.82363 &    9.21  &  4.00  &   0.040 &     10 &$  0.12 $ & \nodata  & $ 0.69 $ & \nodata   &  X/S \\
-- & 0017+296 & J0019+2956 & 00 19 37.791440 & +29 56 01.93471 &   19.34  &  7.63  &  -0.015 &     14 & \nodata  & \nodata  & $ 0.08 $ & \nodata   &  S   \\
-- & 0021-084 & J0024-0811 & 00 24 00.672746 & -08 11 10.05397 &    4.75  &  8.88  &  -0.124 &      7 &$  0.08 $ &$  0.06 $ & $ 0.09 $ & $  0.06 $ &  X/S \\
C  & 0026-015 & J0029-0113 & 00 29 00.986049 & -01 13 41.76047 &    0.74  &  0.95  &   0.349 &     36 &$  0.20 $ &$  0.06 $ & $ 0.19 $ & $  0.10 $ &  X/S \\
C  & 0027-024 & J0030-0211 & 00 30 31.823755 & -02 11 56.13361 &    0.30  &  0.57  &  -0.243 &     68 &$  0.25 $ &$  0.22 $ & $ 0.18 $ & $  0.16 $ &  X/S \\
C  & 0029-147 & J0031-1426 & 00 31 56.411853 & -14 26 19.34670 &    0.93  &  1.60  &  -0.607 &     36 &$  0.17 $ &$ <0.06 $ & $ 0.23 $ & $  0.13 $ &  X/S \\
C  & 0042+186 & J0044+1855 & 00 44 42.227898 & +18 55 05.03460 &    0.76  &  1.69  &   0.591 &     36 &$  0.09 $ & \nodata  & $ 0.14 $ & $  0.11 $ &  X/S \\
-- & 0042-373 & J0045-3705 & 00 45 12.065955 & -37 05 48.47087 &    3.05  &  7.43  &  -0.066 &     20 &$  0.20 $ &$  0.19 $ & \nodata  & \nodata   &  X   \\
C  & 0043+246 & J0046+2456 & 00 46 07.825737 & +24 56 32.52438 &    0.26  &  0.46  &   0.051 &     90 &$  0.44 $ &$  0.33 $ & $ 0.18 $ & $  0.15 $ &  X/S \\
C  & 0046+063 & J0048+0640 & 00 48 58.723150 & +06 40 06.47544 &    0.67  &  1.20  &   0.414 &     28 &$  0.13 $ &$  0.06 $ & $ 0.24 $ & $  0.06 $ &  X/S \\
C  & 0054+161 & J0056+1625 & 00 56 55.294324 & +16 25 13.34088 &    0.33  &  0.29  &   0.019 &     90 &$  0.42 $ &$  0.40 $ & $ 0.27 $ & $  0.26 $ &  X/S \\
C  & 0055+060 & J0058+0620 & 00 58 33.804481 & +06 20 06.07358 &    0.68  &  1.31  &  -0.221 &     19 &$  0.17 $ &$  0.15 $ & $ 0.09 $ & $ <0.06 $ &  X/S \\
\enddata
\tablecomments{Table~\ref{t:cat} is presented in its entirety in the electronic
               edition of the Astronomical Journal. A portion is shown here 
               for guidance regarding its form and contents. Units of right 
               ascension are hours, minutes and seconds, units of declination 
               are degrees, minutes and seconds.}
\end{deluxetable*}

  In total, 365 out of 450 sources were detected and yielded at least 
two good points for position determination. However, not all of these sources
are suitable as phase referencing calibrators or as targets for geodetic 
observations. We consider a source suitable as a calibrator if
1) the number of good X/S pairs of observations is 8 or greater in order to 
rule out the possibility of a group delay ambiguity resolution error; 
and 2) position error before reweighting is less than 5~mas. We found that 
usually 8 observations at the VLBA is sufficient to get positions with 
errors less than 5~mas. If the errors are significantly greater, this means 
that the source was not detected at long baselines, and therefore, we do not 
possess information on whether the source can be detected when used as 
a phase referencing calibrator. Only 231 sources satisfy this calibrator 
criteria. It should be mentioned that our criterion for suitability 
a source as a phase calibrator is rather conservative, and sources which 
fail this criterion may still be useful for some applications.

  Five detected sources turned out to be known gravitational lenses. 
Since the fringe amplitude may have several minima and maxima during 
integration time for such source, our technique for fringe searching 
is not applicable. We excluded these gravitational lenses from the analysis, 
since they are certainly not suitable as calibrators.

\section{The VCS3 catalogue}

  The VCS3 catalogue is listed in Table~\ref{t:cat}. The first column gives 
source class: ``C'' if the source can be used as a calibrator, ``--'' if it 
cannot. The second and third columns give IVS source name (B1950 notation), 
and IAU name (J2000 notation). The fourth and fifth columns give source
coordinates at the J2000.0 epoch. Columns /6/ and /7/ give inflated source
position uncertainties in right ascension and declination in mas, and 
column /8/ gives the correlation coefficient between the errors in right
ascension and declination. The number of group delays used for position 
determination is listed in column /9/. The next two columns give the 
estimates of correlated flux density in Jansky at X-band: total CLEAN flux 
density (column /10/) and flux density of unresolved components (column /11/). 
Similar information for S-band is listed in columns /12/ and /13/. For 
some sources only estimates of the total VLBA flux density are presented, 
because either no data at long baseline were collected or the data at these
baselines were unreliable. Column /14/ gives data type used for position 
estimation: X/S stands for ionosphere-free linear combination of X and 
S wide-band group delays; X stands for X-band only group delays; and 
S stands for S-band only group delays. Some sources which yielded less 
than 8 pairs of X and S band group delay observables had 2 or more 
observations at X and/or S band observations. For these sources either 
X-band or S-band only estimates of coordinates are listed in the VCS3 
catalogue, whichever uncertainty is less.

  In addition to this table, the html version of this catalogue is
posted on the Web at \url{http://gemini.gsfc.nasa.gov/vcs3}. For each source
it has 8 links: to a pair of postscript maps of the source at X and S-band; 
to a pair of plots of correlated flux density as a function of the 
length of the baseline projection to the source plane; to a pair of fits 
files with naturally weighted CLEAN images; and to a pair of fits files with 
calibrated $uv$ data. The coordinates and the plots are also 
accessible from the NRAO VLBA Calibration Search web-page
\url{http://magnolia.nrao.edu/vlba\_calib/}.

\begin{figure*}[h]
   \resizebox{1.0\hsize}{!}
   { 
     \includegraphics[trim=0cm 0cm 0cm 6.0cm]{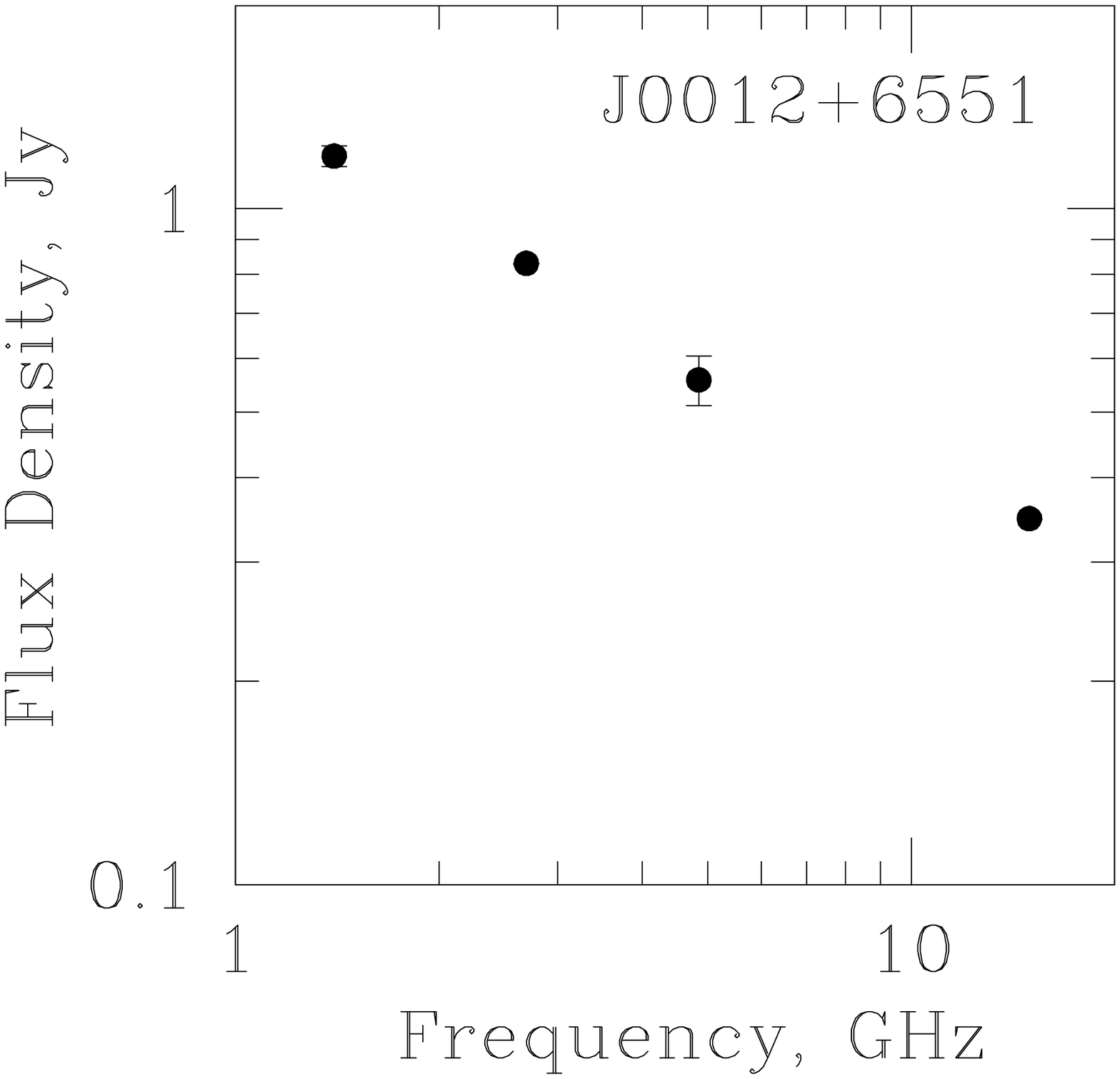}
     \includegraphics[trim=0cm 0cm 0cm 6.0cm]{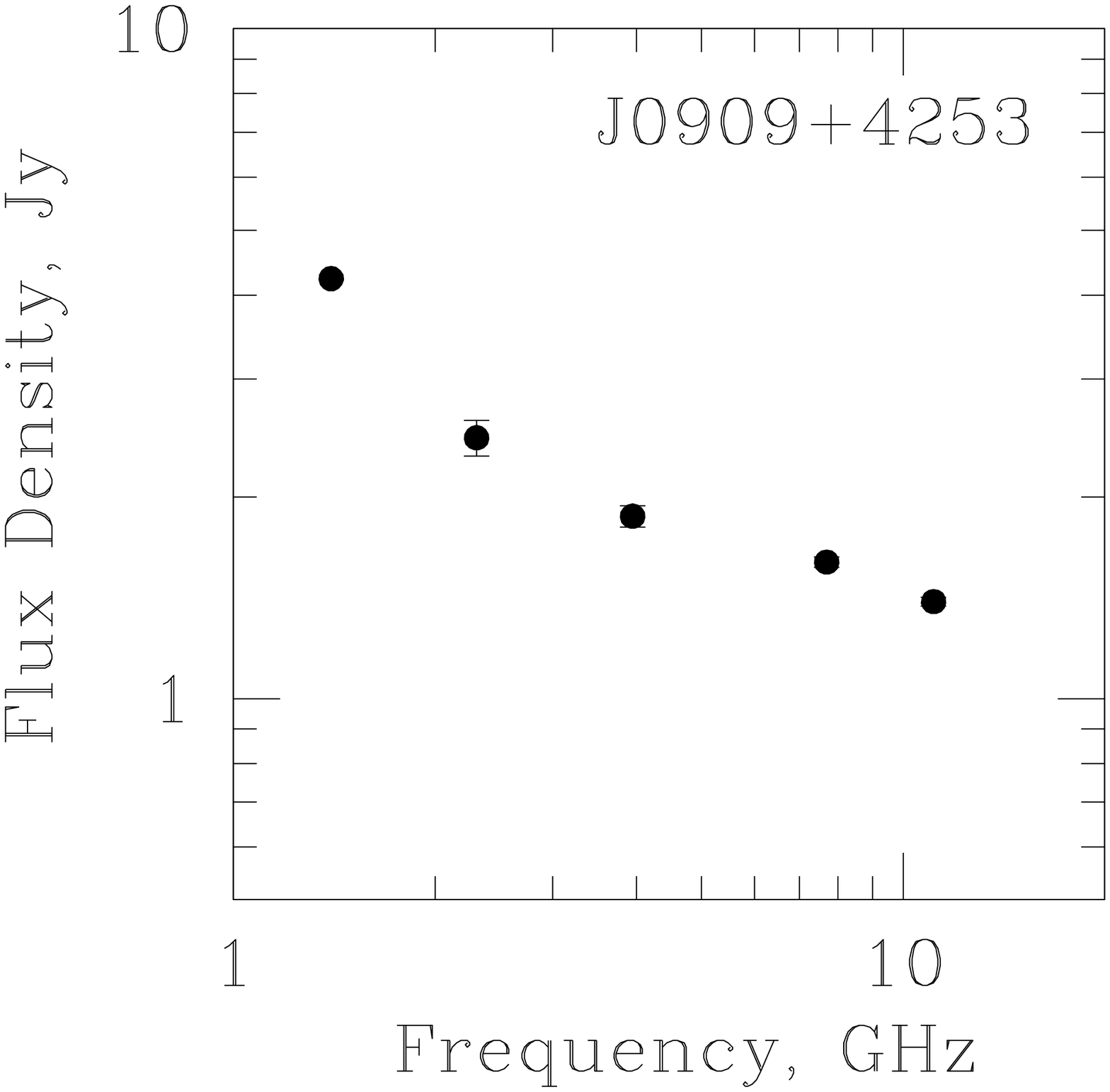}
     \includegraphics[trim=0cm 0cm 0cm 6.0cm]{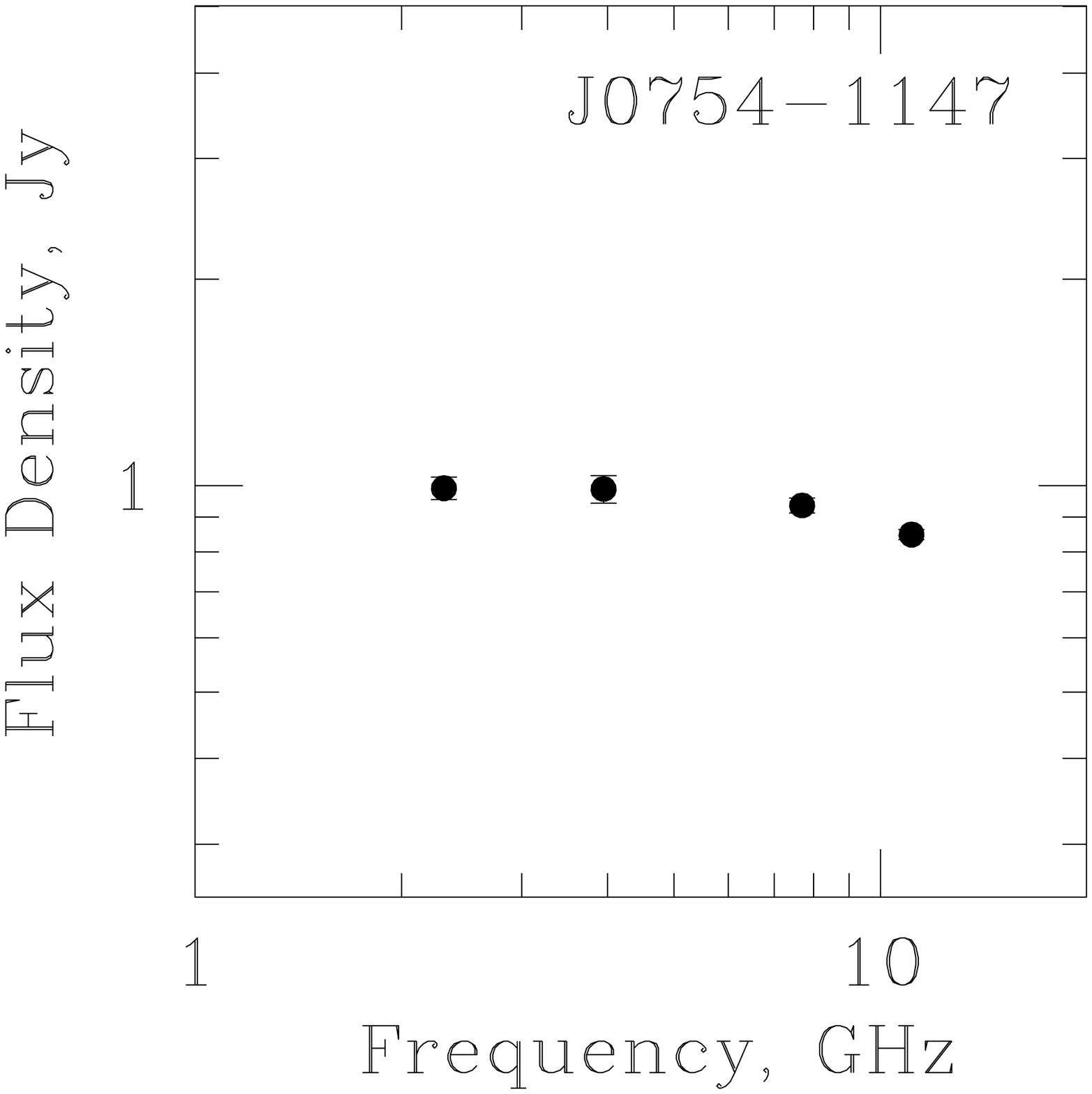}
     \includegraphics[trim=0cm 0cm 0cm 6.0cm]{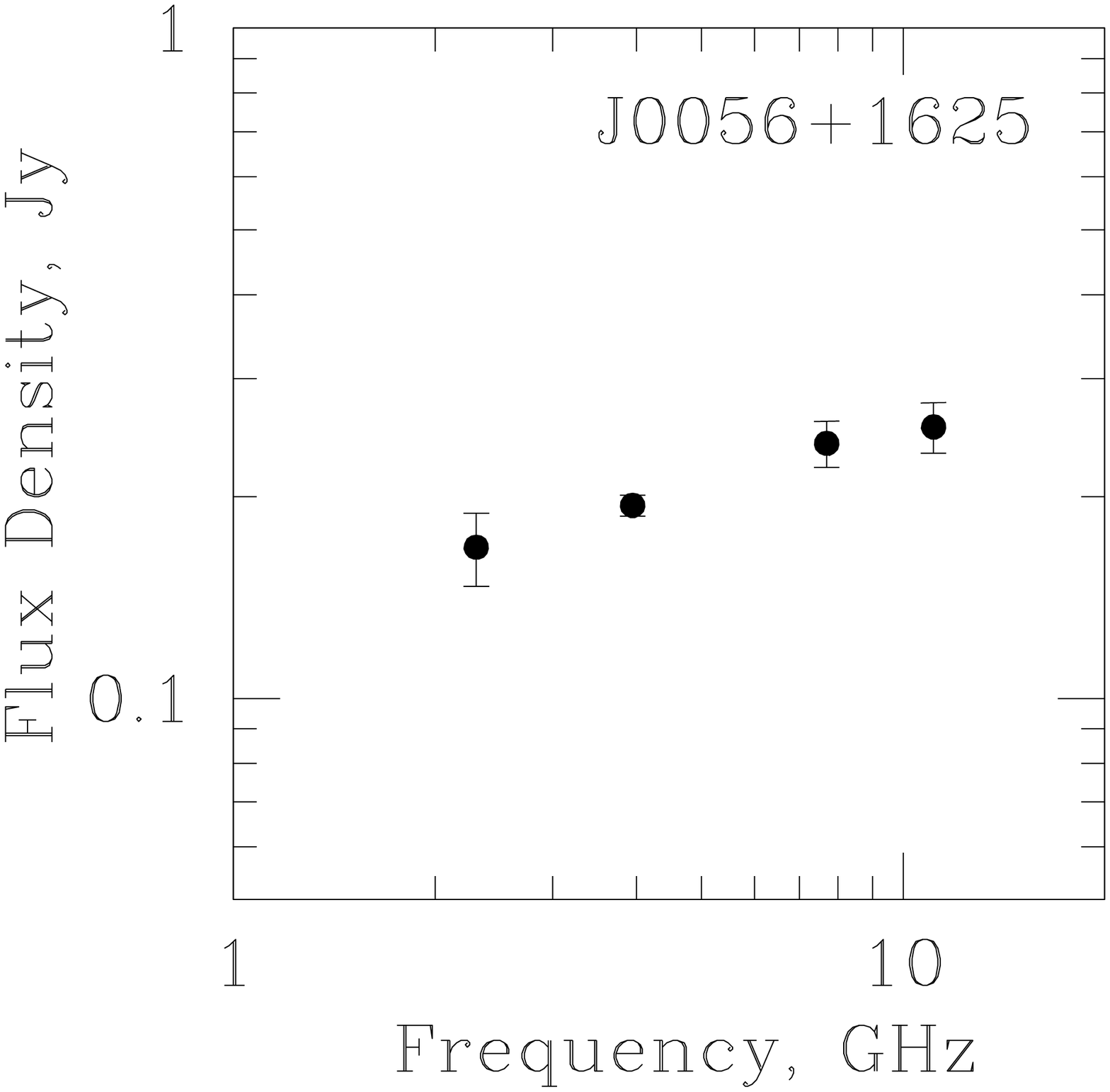}
   }
   \par\vspace{-4.4cm}  a) \par\vspace{4.4cm}
   \resizebox{1.0\hsize}{!}
   { 
     \includegraphics[trim=0cm 1cm 0cm 0cm]{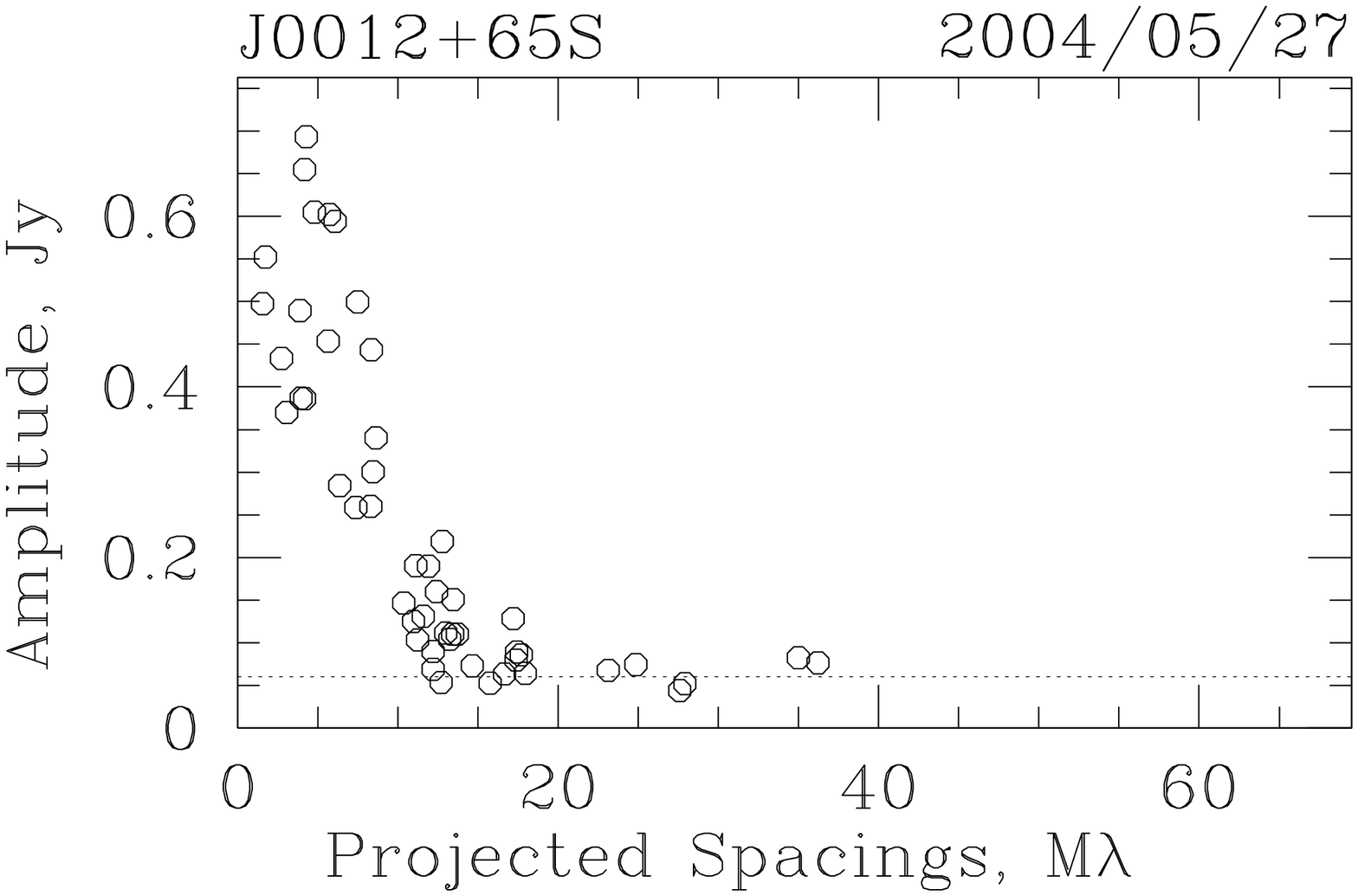}
     \includegraphics[trim=0cm 1cm 0cm 0cm]{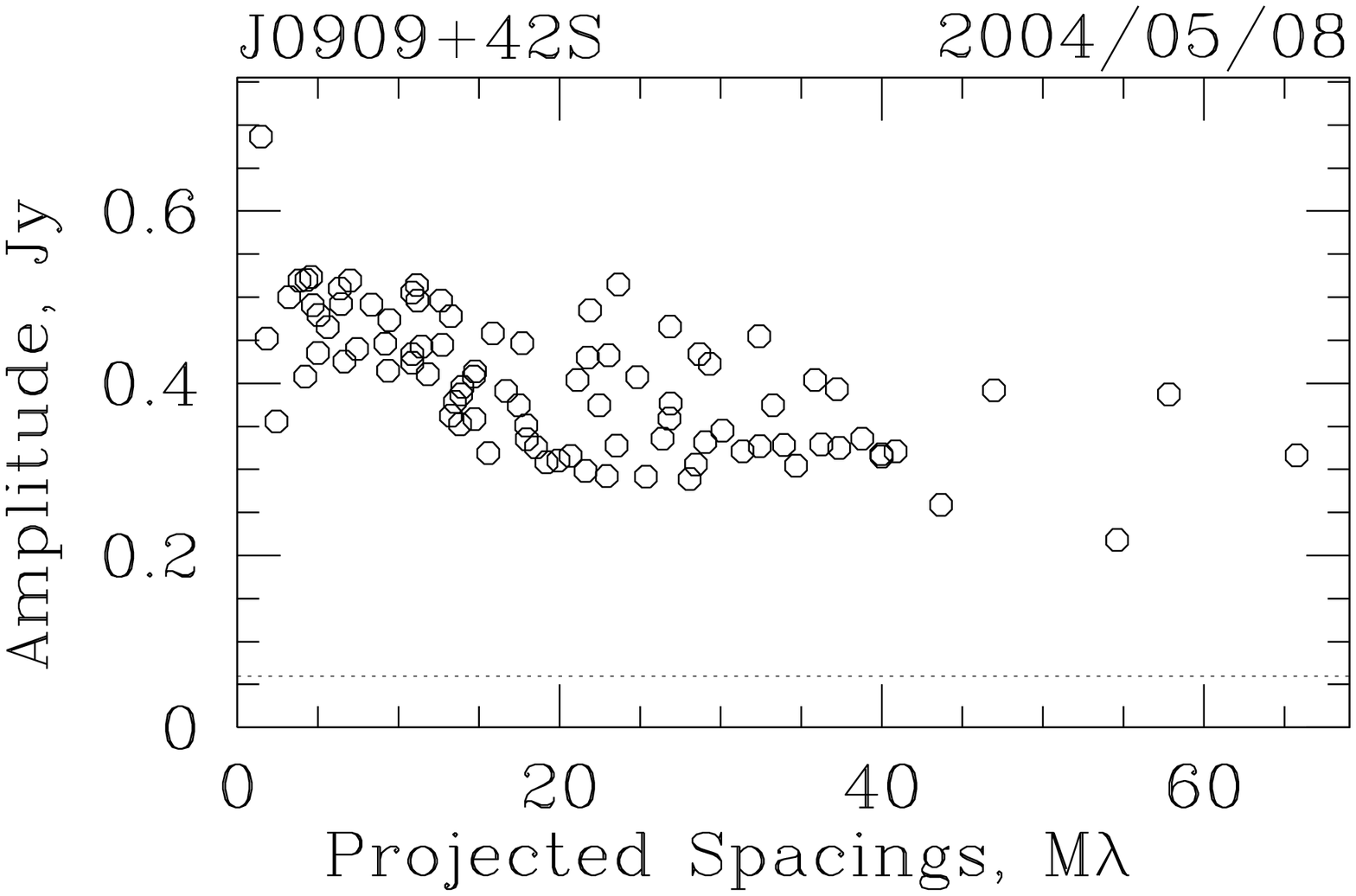}
     \includegraphics[trim=0cm 1cm 0cm 0cm]{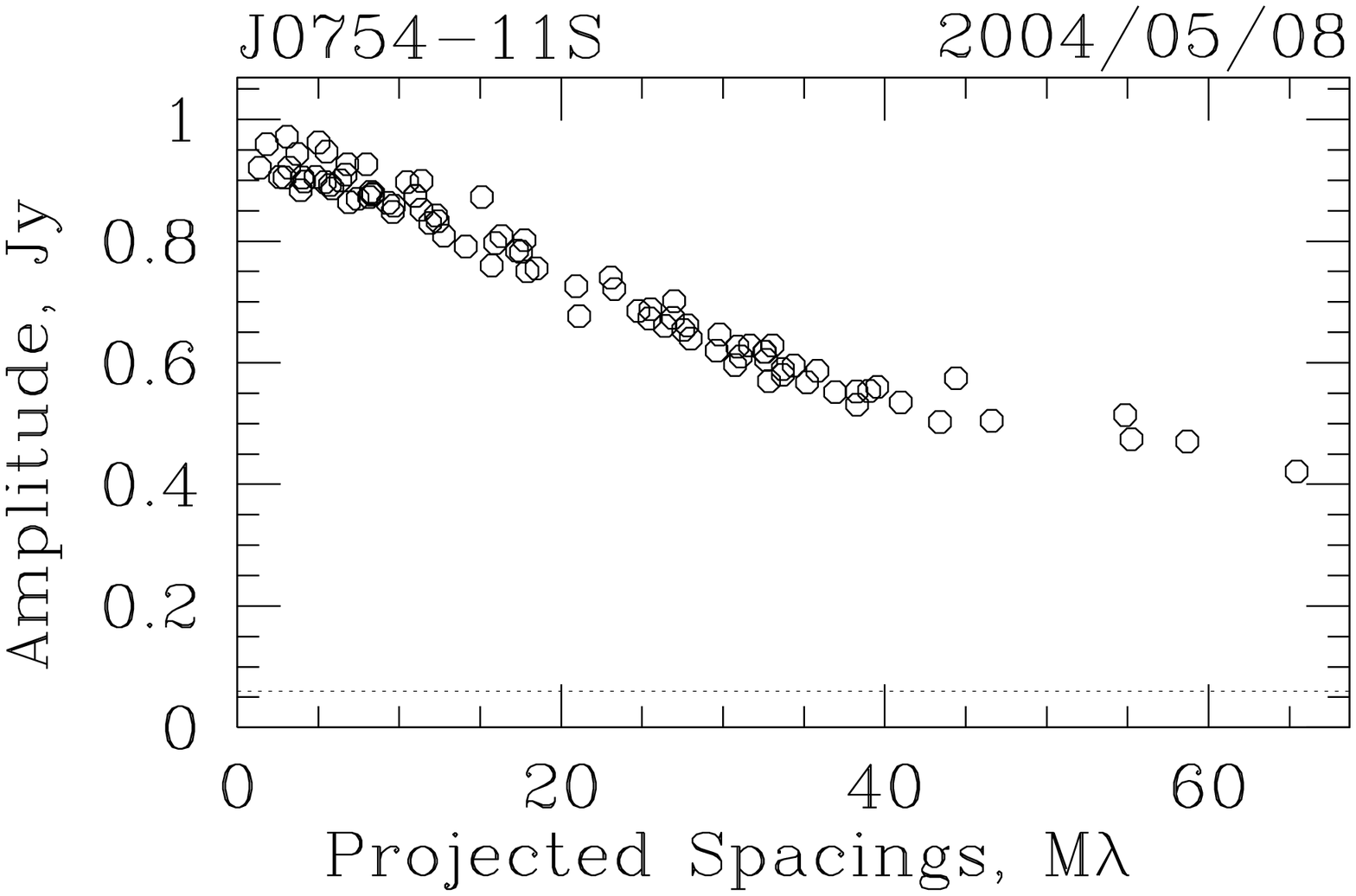}
     \includegraphics[trim=0cm 1cm 0cm 0cm]{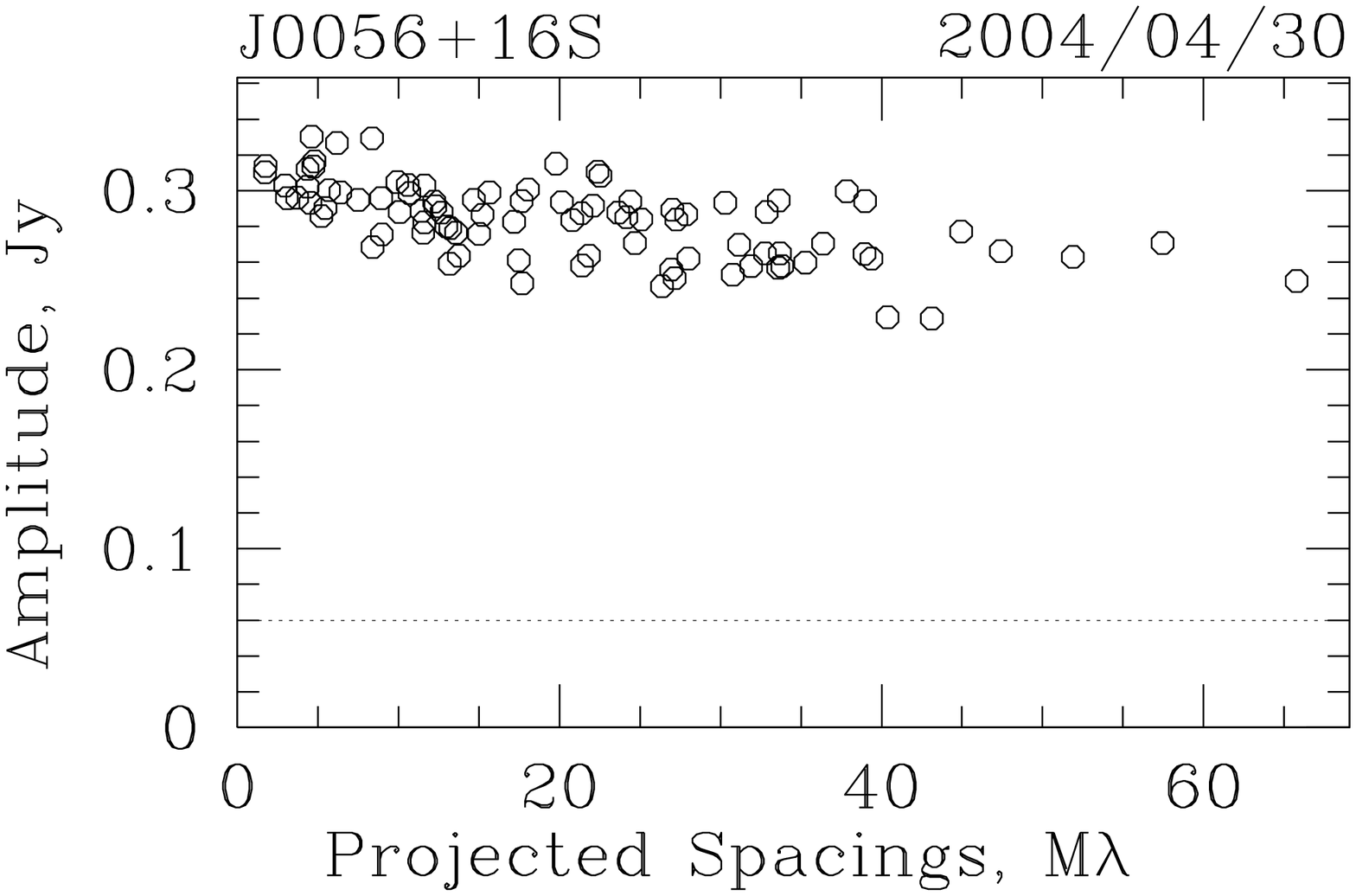}
   }
   \par\vspace{-2.8cm} b) \par \vspace{2.8cm}
   \resizebox{1.0\hsize}{!}
   {
     \includegraphics[trim=1cm 5.4cm 3cm 6.5cm,clip,angle=-90]{Petrov.fig2i.ps}
     \includegraphics[trim=1cm 5.4cm 3cm 6.5cm,clip,angle=-90]{Petrov.fig2j.ps}
     \includegraphics[trim=1cm 5.4cm 3cm 6.5cm,clip,angle=-90]{Petrov.fig2k.ps}
     \includegraphics[trim=1cm 5.4cm 3cm 6.5cm,clip,angle=-90]{Petrov.fig2l.ps}
   }
   \par\vspace{-4.8cm} c) \par \vspace{4.8cm}
   \resizebox{1.0\hsize}{!}
   {
     \includegraphics[trim=0cm 1cm 0cm 0.6cm]{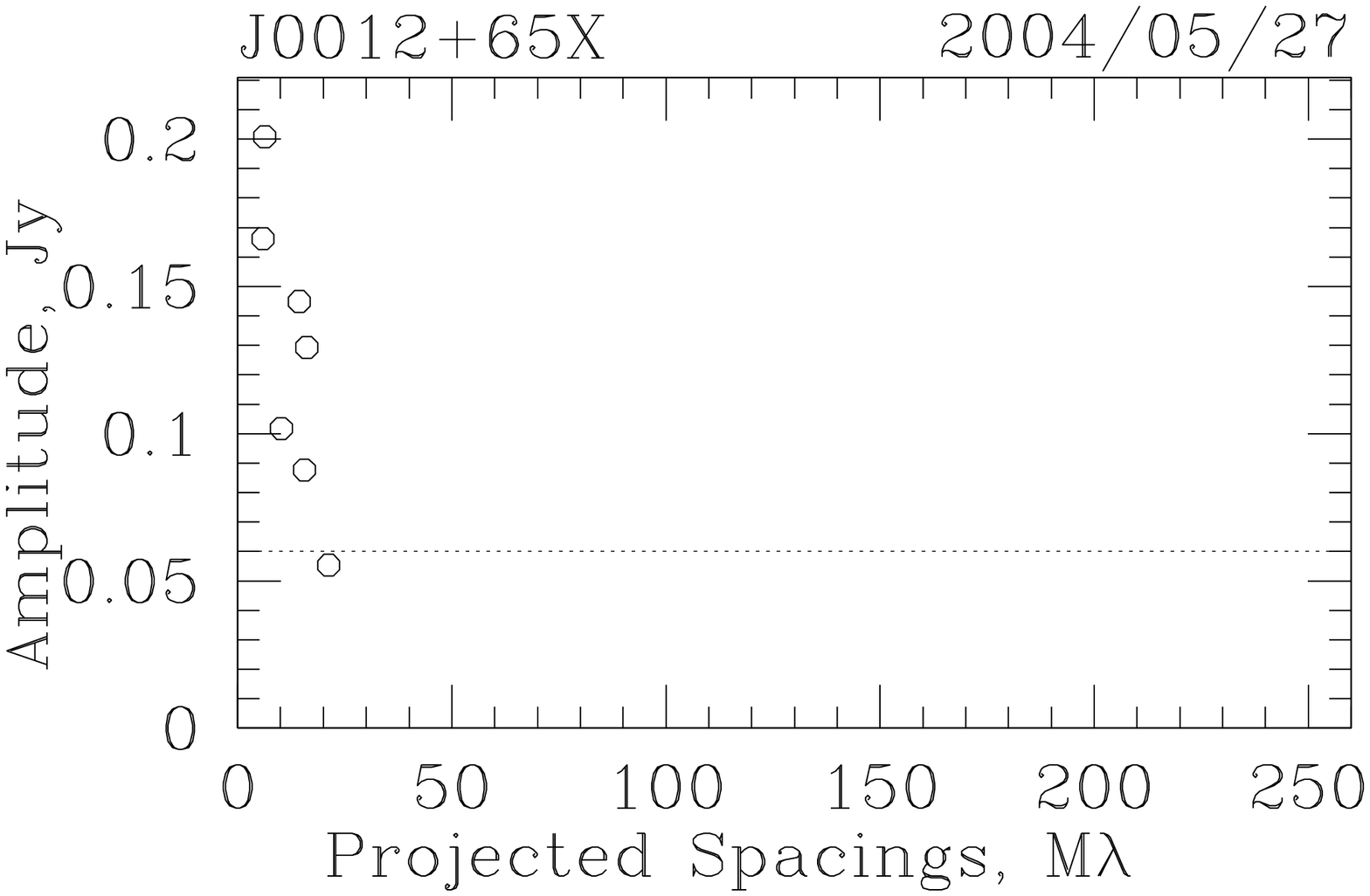}
     \includegraphics[trim=0cm 1cm 0cm 0.6cm]{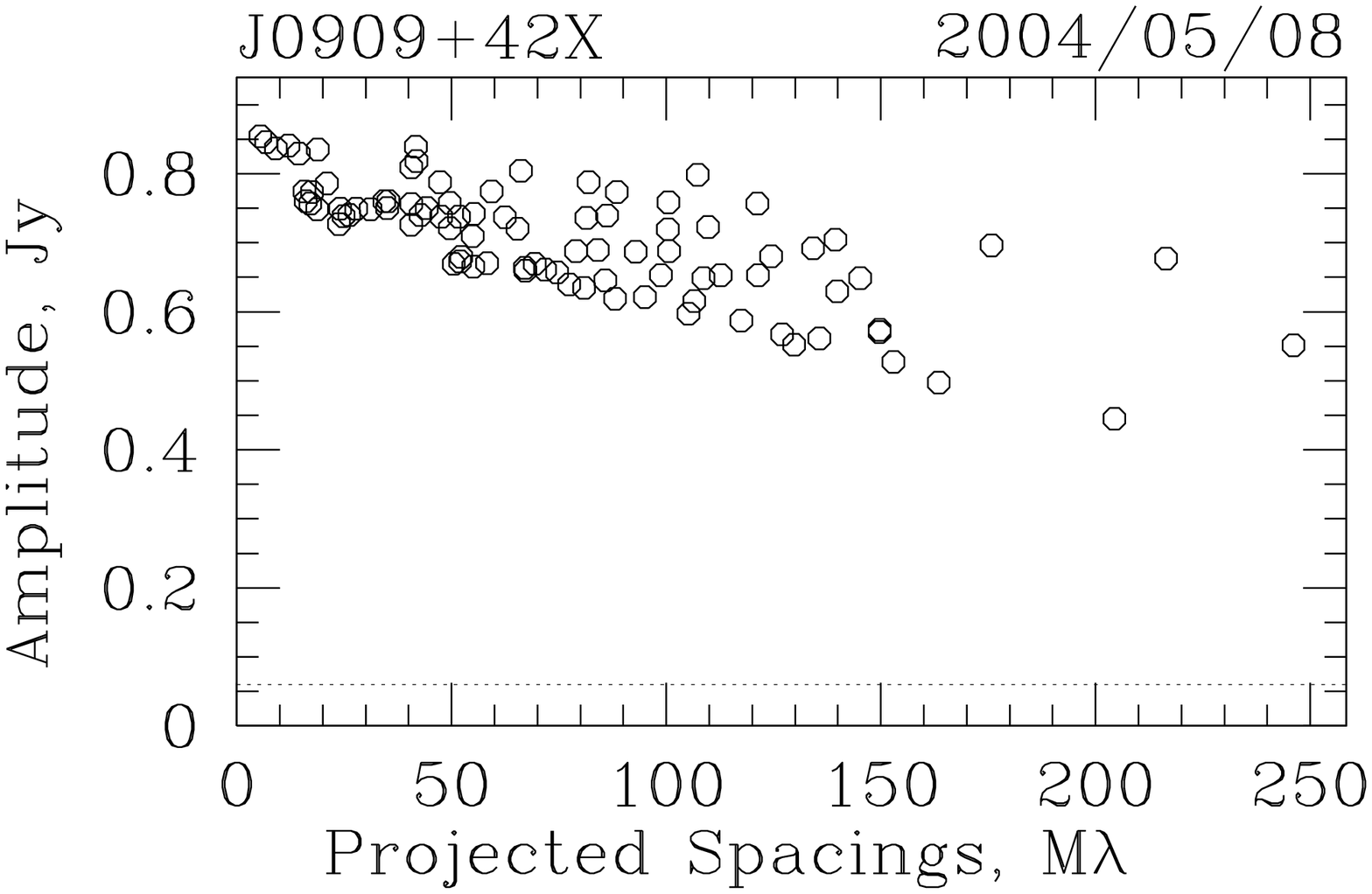}
     \includegraphics[trim=0cm 1cm 0cm 0.6cm]{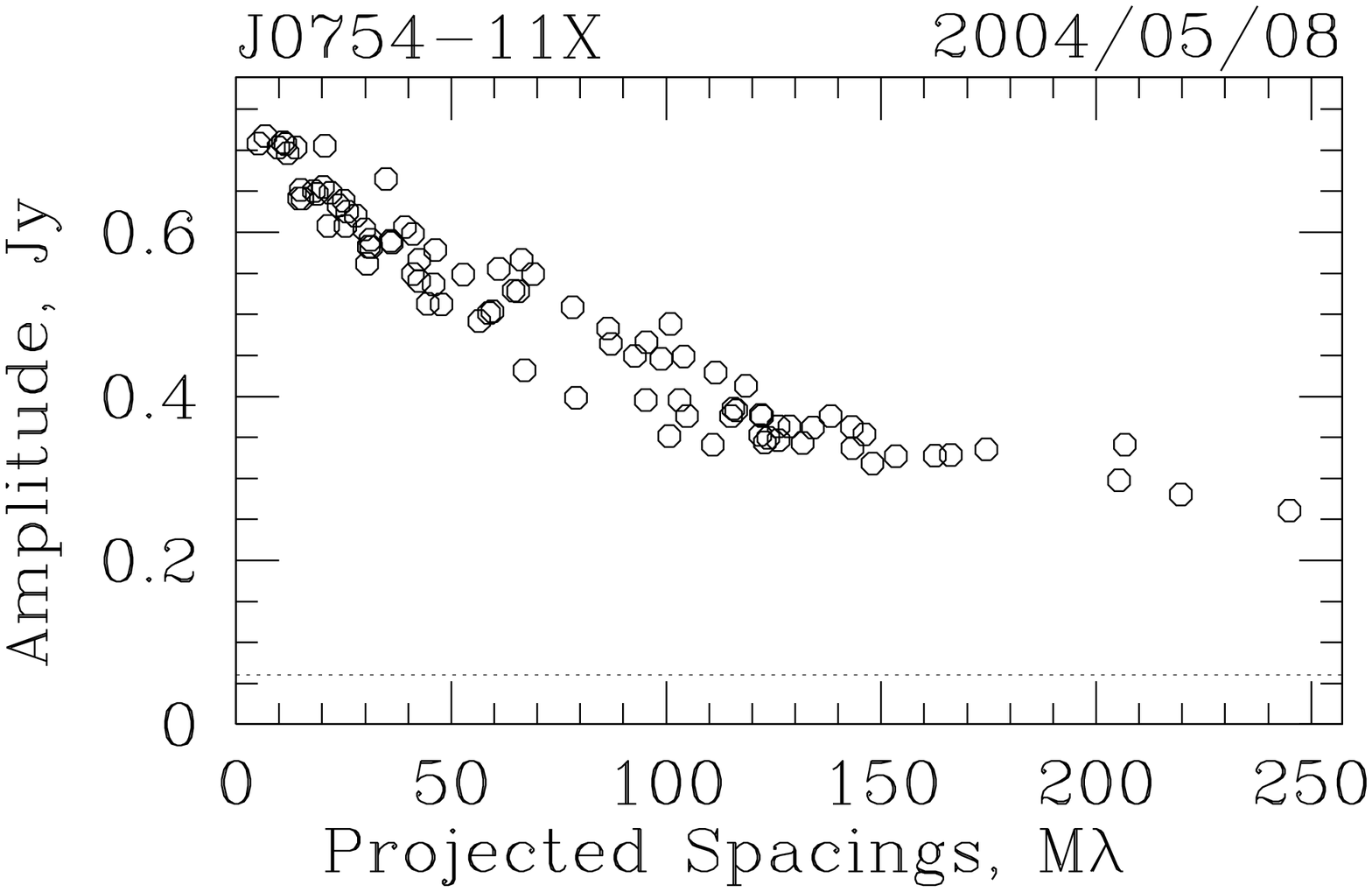}
     \includegraphics[trim=0cm 1cm 0cm 0.6cm]{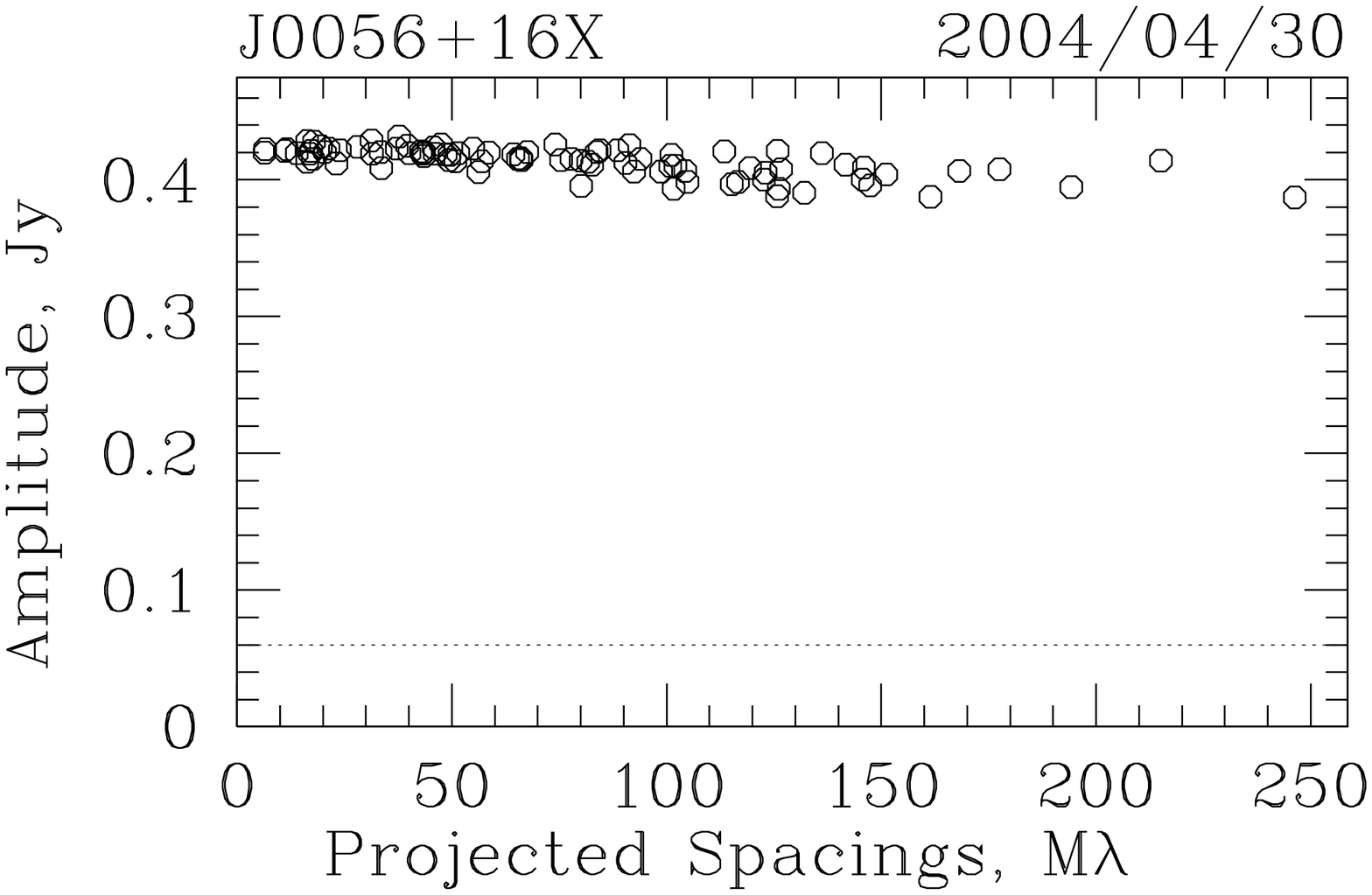}
   }
   \par\vspace{-2.8cm} d) \par\vspace{2.8cm}
   \resizebox{1.0\hsize}{!}
   {
     \includegraphics[trim=1cm 5.4cm 3cm 6.5cm,clip,angle=-90]{Petrov.fig2r.ps}
     \includegraphics[trim=1cm 5.4cm 3cm 6.5cm,clip,angle=-90]{Petrov.fig2s.ps}
     \includegraphics[trim=1cm 5.4cm 3cm 6.5cm,clip,angle=-90]{Petrov.fig2t.ps}
     \includegraphics[trim=1cm 5.4cm 3cm 6.5cm,clip,angle=-90]{Petrov.fig2u.ps}
   }
   \par\vspace{-4.8cm} e) \par\vspace{4.8cm}
\caption{
 a)~Broad-band spectra from \mbox{RATAN--600} observations in 2003 plus data 
   from CATS.
 b)~Dependence of the correlated flux density at S band (2.3~GHz) versus 
    projected spacings. A dotted line shows the detection limit.
 c)~Naturally weighted CLEAN images at S band. The lowest contour levels 
    (2 steps) on images are plotted at ``clev'' levels [Jy/beam], the peak 
    brightness~--- ``max'' values [Jy/beam]. The dashed contours indicate 
    negative flux. The beam is shown in the bottom left corner of the images.
 d)~Dependence of the correlated flux density at X band (8.6~GHz) versus 
    projected spacings.
 e)~Naturally weighted CLEAN images at X band. 
 \label{f:images}
}
\end{figure*}

\begin{figure*}[t]
   \begin{center}
     \resizebox{1.0\hsize}{!}
        {
          \includegraphics[trim=0cm 0cm 0cm 0cm,clip,angle=0]
                           {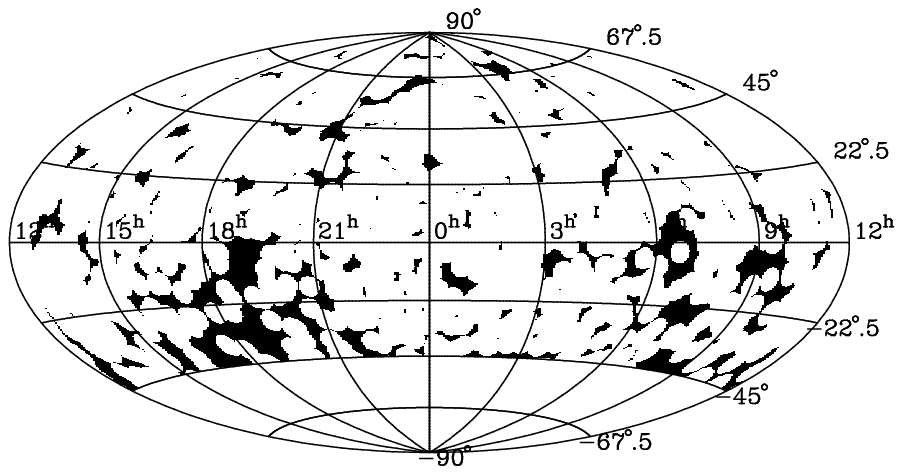}
          \includegraphics[trim=-1.0cm 0cm 0cm 0cm,clip,angle=0]
                           {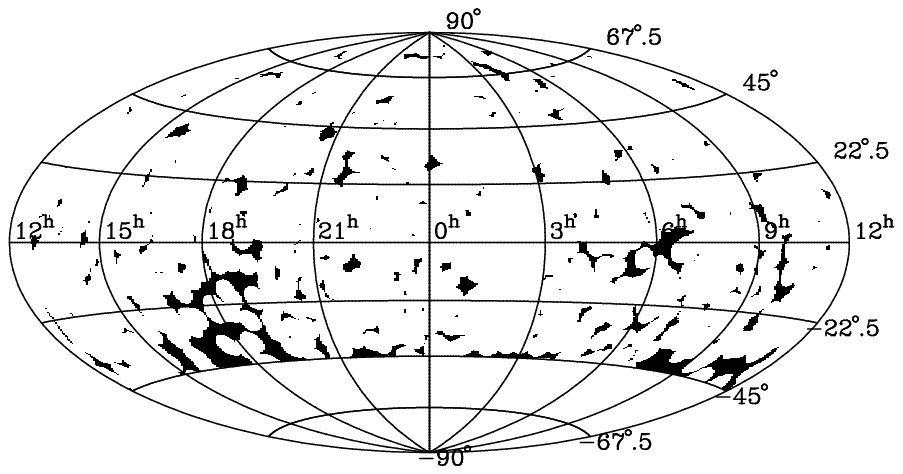}
        }
   \end{center}
   \caption{Source sky density before (left) and after (right) the VCS3 
            observing campaign. The area above $-45\degr\ $ declination 
            with calibrator source density less than one source within 
            a 4\degr\ radius disk is shown in black. 
            \label{f:skymap}
           }
\end{figure*}

  A sample of data used for candidate selection as well as some typical
imaging results are shown in Figure~\ref{f:images}. In row /a/ we present the 
broad-band spectra from \mbox{RATAN--600} observations in 2003 and from some 
old data taken from the NVSS, 87GB \citep{87gb} and the Radio Survey of the
Galactic Plane at 11cm \citep{11cmb} catalogues. In rows /b/ and /d/ 
the dependence of visibility function amplitude versus projected 
baseline at S and X bands, respectively, is shown. Each point is a coherent 
average over the scan duration and over the band. Naturally weighted CLEAN 
images at S and X bands are presented in rows /c/ and /e/\footnote{The scale 
of the J0012+6551 map is increased with respect to other maps.}. 
The sources in this figure are placed such that the contribution 
of the compact component in total emission increases from left to right. 
According to \citet{kov02}, we interpret the broad-band spectra of all 
selected sources as a sum of two constituents: emission from an optically 
thin extended component with steep spectrum which usually increases with 
decreasing frequency up to hundreds of MHz, and emission from a compact 
component with self-absorption which has a maximum at a frequency higher 
than 1~GHz. The spectrum of J0012+6551 has spectral index $-0.5$. 
No contribution to emission from the compact component is seen, and the 
correlated flux density even at the shortest baseline projections is several 
times less than the total flux density. See discussion of angular dimensions 
of compact steep spectrum sources in \citet{css_gps} for more details. 
The source is below the detection limit at both bands at spacings 
\mbox{$>$ 20~M$\lambda$}. At X-band the map shows only noise. The spectrum 
of J0909+4253 has the spectral index $-1.1$ at frequencies less than 2~GHz 
and it flattens at higher frequencies. This flattening is caused by the 
contribution of the compact component. When the contribution of the extended 
component is subtracted from the spectrum, the residual spectrum grows with 
increasing the frequency, and our measurements of the correlated flux density 
confirm it. The remaining two sources have flat and inverted spectra, 
which is an indication of dominance of the compact constituent in total 
emission. Again our VLBA observations confirm it. Correlated flux densities
at short spacings for J0056+1625 and some other sources are even higher than 
the total flux density observed at RATAN half a year before the VLBA 
measurements due to the strong variability of the object. Analysis of this 
sample illustrates the validity of our approach for candidate selection.

\section{Conclusion}

   The VCS3 Survey has added 360 new sources not previously observed
with VLBI. Of these, 231 sources are suitable as phase referencing calibrators
and as target sources for geodetic applications. A careful strategy of
source selection resulted in a very high detection rate; more than 
80\% of the sources in our candidates list were detected.

   This campaign contributed considerably in reducing the areas with low density
of known calibrators. If only sources from the ICRF-Ext2 catalogue are taken 
into account, 48.1\% of the sky area above declination $ -45\degr\ $ has no 
calibrator within a 4\degr\ radius disk in any given direction. Combining the
ICRF-Ext2 and the VCS1 catalogues, this area is reduced to 13.7\%. Adding the 
VCS2 catalogue reduces the area to 10.5\%. After the VCS3 campaign this area 
is only 5.3\% of the sky above declination $-45\degr$. Figure~\ref{f:skymap}
shows the area with insufficient calibrators density before and after the 
VCS3 observing campaign.

\acknowledgments
The VLBA is a facility of the National Radio Astronomy Observatory
which is operated by Associated Universities, Inc., under a cooperative
agreement with the National Science Foundation. The authors made use of
the database CATS \citep{cats} of the Special Astrophysical Observatory.
This research has made use of the NASA/IPAC Extragalactic Database
(NED) which is operated by the Jet Propulsion Laboratory, California
Institute of Technology, under contract with the National Aeronautics
and Space Administration. \mbox{RATAN--600} observations were partly 
supported by the Russian State Program ``Astronomy'' and the Russian 
Ministry of Education and Science, the NASA JURRISS Program 
(project W--19611), and the Russian Foundation for Basic Research 
(projects 01--02--16812 and 02--02--16305). The authors are thankful 
to D.~McMillan for valuable comments. This work was done while L.~Petrov 
and D.~Gordon worked for NVI, Inc. and Raytheon, respectively, under NASA 
contract NAS5--01127.


\begin{thebibliography}{99}

   \bibitem[Beasley et~al.(2002)]{vcs1} Beasley,~A.~J., Gordon,~D.,
            Peck,~A.~B., Petrov,~L., MacMillan,~D.~S., Fomalont,~E.~B., \&
            Ma,~C.\
	    2002, \apjs, 141, 13

   \bibitem[Bolton, Savage, \& Wright(1971)]{pkscat90}
            Bolton,~J.~G., Savage,~A., \& Wright,~A.~E.\
            1979, Australian Journal of Physics, Astroph.~Suppl., 46, 1

   \bibitem[Cohen \& Shaffer(1971)]{first-cat}
            Cohen,~M.~H., \& Shaffer,~D.~B.\
	    1971, \aj, 76, 91.

   \bibitem[Condon et~al.(1998)]{nvss}
           Condon,~J.~J., Cotton,~W.~D., Greisen,~E.~W., Yin,~Q.~F., 
           Perley,~R.~A., Taylor,~G.~B., \& Broderick,~J.~J.\
	   1998, \aj, 115, 1693

   \bibitem[Fey et~al.(2004)]{icrf-ext2-2004} Fey,~A.L., Ma,~C., 
            Arias,~E.~F., Charlot,~P., Feissel-Vernier,~M., Gontier,~A.-M.,
            Jacobs,~C.~S., Li,~J., \& MacMillan,~D.~S.\
	    2004, \aj, 127, 3587.

   \bibitem[Fomalont et~al.(2003)]{vcs2} 
            Fomalont,~E., Petrov,~L., McMillan,~D.~S., Gordon,~D., \& Ma,~C.\ 
            2003, \aj, 126, 2562

   \bibitem[F\"urst et al.(1990)]{11cmb}
            F\"urst,~E., Reich,~W., Reich,~P., \& Reif,~K.\
	    1990, \aaps, 85, 805

   \bibitem[Gregory \& Condon(1991)]{87gb}
            Gregory,~P.~C., \& Condon,~J.~J.\
	    1991, \apjs, 75, 1011

   \bibitem[Gregory et~al.(1996)]{gb6}
            Gregory,~P.~C., Scott,~W.~K., Douglas,~K., \& Condon,~J.~J.\
            1996, \apjs, 103, 427

   \bibitem[Greisen(1988)]{aips}
            Greisen,~E.~W.\
	    1988, in Acquisition,
            Processing and Archiving of Astronomical Images, ed. G.~Longo \&
            G.~Sedmak (Napoli: Osservatorio Astronomico di Capodimonte), 125


   \bibitem[Jackson et~al.(2002)]{parkes-qJy}
            Jackson,~C.~A., Wall,~J.~V., Shaver,~P.~A., Kellermann,~K.~I.,
            Hook,~I.~M., \& Hawkins,~M.~R.~S.\
	    2002, \aap, 386, 97

   \bibitem[Kellermann, Pauliny-Toth, \& Davis(1968)]{kellermann}
          Kellermann,~K.~I., Pauliny-Toth,~I.~I.~K., \& Davis,~M.~M.\
          1968, Astrophysical Letters, 2, 105

   \bibitem[Kellermann \& Owen(1988)]{ko88}
          Kellermann,~K.~I. \& Owen,~F.~N.\
	  1988, Galactic and Extragalactic 
          Radio Astronomy, ed.\ G.~L.~Verschur \& K.~I.~Kellermann
          (Springer Verlag), 563 

   \bibitem[Kovalev et~al.(1999)]{Kovalev_etal99}
            Kovalev,~Y.~Y., Nizhelsky,~N.~A., Kovalev,~Yu.~A., Berlin,~A.~B.,
            Zhekanis,~G.~V., Mingaliev,~M.~G., \& Bogdantsov,~A.~V.\
            1999, \aaps, 139, 545

   \bibitem[Kovalev et~al.(2002)]{kov02}
            Kovalev,~Y.~Y., Kovalev,~Yu.~A., Nizhelsky,~N.~A., \& 
            Bogdantsov,~A.~B.\
	    2002, PASA, 19, 83

   \bibitem[Ma et~al.(1998)]{icrf98}
            Ma,~C., et~al.\
            1998, \aj, 116, 516

   \bibitem[Matveenko et~al.(1965)]{mat65}
            Matveenko,~L.~I., Kardashev,~N.~S., Sholomitskij,~G.~B.\
	    1965, Izvestia VUZov.\ Radiofizika, 8, 651 (in Russian)

   \bibitem[Myers et~al.(2003)]{class}
            Myers,~S.~T., et~al.\
	    2003, \mnras, 341, 1

   \bibitem[O'Dea(1998)]{css_gps}
        O'Dea, C.~P.\
	1998, \pasp, 110, 493

   \bibitem[Pearson et al.(1994)]{difmap-script}
            Pearson,~T.~J., Shepherd,~M.~C., Taylor,~G.~B., \& Myers,~S.~T.\
            1994, \baas, 185, 0808

   \bibitem[Ryan et~al.(1993)]{errfactor}
          Ryan,~J.~W., Clark,~T.~A., Ma,~C., Gordon,~D. Caprette,~D.~S., \& 
          Himwhich,~W.~E.\
	  1993, in Global scale tectonic plate motions measured 
          by CDP VLBI data, 23, ed.\ D.~E.~Smith \& D.~L.~Turcotte,
          (Washington, D.C.: American Geophysical Union, Geodynamics 
          Series), 37

  \bibitem[Shepherd(1997)]{difmap}
          Shepherd,~M.~C.\
	  1997, in ASP Conf.\ Series.\ 125,
          Astronomical Data Analysis Software and Systems~VI,
          ed.\ G.~Hunt \& H.~E.~Payne (San Francisco: ASP), 77

   \bibitem[Verkhodanov et~al.(1997)]{cats}
            Verkhodanov,~O.~V., Trushkin,~S.~A., Andernach,~H., \& 
            Chernenkov,~V.~N.\
	    1997, in ASP Conf.\ Ser.\ 125, 
            Astronomical Data Analysis Software and Systems VI, 
            ed.\ G.~Hunt, \& H.~E.~Payne (San Francisco: ASP), 322

   \bibitem[Winn, Patnaik, \& Wrobel(2003)]{phcal-south}
	    Winn,~J.~N., Patnaik,~A.~R., \& Wrobel,~J.~M.\
	    2003, \apjs, 145, 83

   \bibitem[Wright and Otrupcek(1990)]{pkscat90a}
            Wright,~A.~E. \& Otrupcek,~R.\
	    1990, Australia Telescope National Facility, CSIRO, Parkes

   \bibitem[Wright et~al.(1996)]{pmn}
            Wright,~A.~E., Griffith,~M.~R., Hunt,~A.~J., Troup,~E., 
            Burke,~B.~F., \& Ekers,~R.~D.\
	    1996, \apjs, 103, 145

   \bibitem[Wrobel et~al.(1998)]{jvas}
	    Wrobel,~J.~M., Patnaik,~A.~R., Browne,~I.~W.~A., \& Wilkinson,~P.~N.\
            1998, \aaps, 193, 4004

   \bibitem[Wrobel and Ulvestad(2004)]{VLBA_summ}
            Wrobel,~J.~M., \& Ulvestad,~J.~S.\
	    2004, VLBA status summary, NRAO

\end{thebibliography}
\end{document}